\documentstyle[aps,prl,amssymb,latexsym]{revtex}

\newcommand{\beq}{\begin{equation}}
\newcommand{\eeq}{\end{equation}}
\newcommand{\bqa}{\begin{eqnarray}}
\newcommand{\eqa}{\end{eqnarray}}
\newcommand{\fr}{\frac}

\begin{document}

\draft

\wideabs{
\title{Integrability of the Minimal Strain Equations for the Lapse and Shift in $3+1$  Numerical Relativity}
\author{S\' ergio M. C. V. Gon\c calves}
\address{Theoretical Astrophysics, California Institute of Technology, Pasadena, California 91125}
\date{February 23, 2000}
\maketitle
\begin{abstract}
Brady, Creighton and Thorne have argued that, in numerical relativity simulations of the inspiral of binary black holes, if one uses lapse and shift functions satisfying the ``minimal strain equations'' (MSE), then the coordinates might be kept co-rotating, the metric components would then evolve on the very slow inspiral timescale, and the computational demands would thus be far smaller than for more conventional slicing choices. In this paper, we derive simple, testable criteria for the MSE to be strongly elliptic, thereby guaranteeing the existence and uniqueness of the solution to the Dirichlet boundary value problem. We show that these criteria are satisfied in a test-bed metric for inspiraling binaries, and we argue that they should be satisfied quite generally for inspiraling binaries. If the local existence and uniqueness that we have proved holds globally, then, for appropriate boundary values, the solution of the MSE exhibited by Brady et. al. (which tracks the inspiral and keeps the metric evolving slowly) will be the unique solution and thus should be reproduced by (sufficiently accurate and stable) numerical integrations.
\end{abstract}
}

\narrowtext

\section{INTRODUCTION}

Brady, Creighton and Thorne \cite{bct98} have shown that post-Newtonian techniques are unable to treat the late stages of inspiral of black-hole binaries with sufficient accuracy to satisfy the needs of gravitational-wave data analysis. The only viable way to perform these computations appears to be numerical relativity. The challenge of performing such computations is called the Intermediate Binary Black Hole (IBBH) problem: the late inspiral is intermediate between the early inspiral (where post-Newtonian techniques {\em are} successful) and the merger (where numerical relativity {\em must} be used). No researchers have yet tackled the IBBH problem by numerical relativity, but it is a high-priority, near-future challenge \cite{bct98}.

If conventional numerical relativity techniques are used for the IBBH problem, then the coordinate system will be asymptotically inertial and the black holes will move through it dynamically on near circular orbits. As a result, the typical timescale for strong evolution of the metric at generic locations will be set by the orbital angular frequency $\Omega$, $\tau_{\rm orb}\sim\Omega^{-1}$, and near each hole's horizon it will be much smaller than this. To follow these evolutions stably will require very short timesteps, $\Delta t\ll\tau_{\rm orb}$, which can be computationally very expensive.

A promising alternative, proposed by Brady et. al., is to use co-rotating coordinates and a wisely chosen time coordinate, in which the metric components evolve (everywhere, including near the horizons) on the gravitational radiation reaction timescale, $\tau_{\rm rr}\gg\tau_{\rm orb}$. In such coordinates, the timesteps can be much longer than in the conventional coordinates, thereby reducing the computation time by a large factor.

Brady et. al. \cite{bct98} have proposed a specific choice for the lapse and shift functions $\alpha$ and $\beta_{j}$ of numerical relativity that, they argue, will force the coordinates to remain co-rotating and the metric components to evolve on the long timescale $\tau_{\rm rr}$. This choice is one that minimizes, at each time step, the integral of the (positive definite) square of the time derivative of the 3-metric, $\int \dot{\gamma}_{ij}\dot{\gamma}_{kl}\gamma^{ik}\gamma^{jl}{\rm d}^{3}\Sigma$, over the computation's spatial coordinate grid. This minimization leads to a set of {\em minimal strain equations} (MSE) \cite{bct98} for the lapse and shift:
\bqa 
\alpha = \fr{K^{ij}D_{i}\beta_{j}}{K^{mn}K_{mn}}, \label{lapse} \\
D^{j}[-\alpha K_{ij}+D_{(i}\beta_{j)}] = 0, \label{shift} 
\eqa
where $D_{i}$ is the 3-covariant derivative. Substitution of
(\ref{lapse}) in (\ref{shift}) yields a second-order
linear differential equation for $\beta_{k}$:
\beq
M_i^{\; k}[D,D]\beta_k = S_i , \label{mse}
\eeq
where the differential operator is
\beq
M_i^{\; k}[D,D] \equiv -\chi^{-1}K_{ij}K^{lk}D^{j}D_{l}+\fr{1}{2}\left(
D_{i}D^{k}+\delta_{i}^{k}D_{j}D^{j} \right), \label{mse2}
\eeq
and
\bqa
S_{i} & \equiv & D_{m}\beta_{n}D^{j}\left(\chi^{-1}K^{mn}K_{ij}\right), \\
\chi & \equiv & K^{ij}K_{ij}.
\eqa

The purpose of this paper is to show that the system (\ref{mse}) is strongly elliptic and thus admits a unique analytical solution for given Dirichlet boundary data, on a sufficiently small compact set. If our results can be extended to global existence and uniqueness, then the results will confirm the MSE as an alternative tool for numerical solutions to the IBBH problem.

The paper is organized as follows: Section 2 presents basic notions and results from regularity theory for elliptic systems of partial differential equations. Section 3 derives necessary and sufficient criteria for the strong ellipticity of the MSE and presents a step-by-step algorithm that can be used to test strong ellipticity for an arbitrary slicing of an arbitrary spacetime metric. In Sec. 4 we use our algorithm to verify strong ellipticity for the optimal slicing of the Thorne test-bed metric \cite{thorne98} (which is an approximation to that of a relativistic, inspiraling binary star), and we then argue that for any inspiraling binary the MSE are likely to be strongly elliptic. The nature of the boundary data is briefly addressed. Section 5 concludes with a summary and comments on future numerical relativity schemes using the MSE.

\section{ELLIPTIC SYSTEMS OF PARTIAL DIFFERENTIAL EQUATIONS}

First, we introduce the notion of strong ellipticity, and then briefly discuss the integrability of strongly\footnote{There are 
weaker notions of ellipticity that still guarantee the {\em existence} of a {\em weak} (defined in a distributional sense) solution, but they may not correspond to a
{\em unique analytical} solution. Strong ellipticity avoids this ambiguity.} elliptic systems.

Consider the following second-order differential equation for a vector field $u^{k}$ defined
on some $n-$dimensional manifold ${\cal M}$:
\beq
M^{i}_{k} [D, D]u^{k}=S^{i}[D],
\eeq
where $M^{i}_{k}[D,D]$ includes the second-order derivative terms and all of the remaining terms are
collected in the `source' vector functional $S^{i}$. The above equation is said to be strongly elliptic if, upon formal substitution at each point on ${\cal M}$ of $D_{j}$ by a vector field $\xi_{j}$, the matrix thereby obtained, $M^{i}_{k}[\vec{\xi},\vec{\xi}]$, has the following property: the
eigenvalues are all positive and bounded away from zero for all $\xi_{j}$ and all points in the manifold ${\cal M}$ (see e.g.~\cite{giaquinta83}).

The Dirichlet problem for a vector field with analytical components $u^{k}(\mbox{\bf x}): \Omega\subset {\Bbb{R}}^{n}\mapsto \Bbb{R}$ can be formally stated as
\bqa
Lu^{k} & = & 0 \; , \; \mbox{\bf x}\, \in\, \Omega, \label{diri1} \\
u^{k} & = & f^{k} \; , \; \mbox{\bf x}\, \in\, \partial\Omega, \label{diri2}
\eqa
where $L$ is an elliptic differential operator and $f^{k}$ are analytical components of a vector field $\vec{f}$ defined on the boundary $\partial\Omega$. The Neumann
problem consists in specifying the first derivative of $u^{k}$ normal to $\partial\Omega$, rather than $u^{k}$ itself. We shall be concerned with the Dirichlet problem for the remainder of this paper, since our boundary data will consist solely of $\beta_{k}$ itself. (see, however, subsec. 4. B).

It can be shown (see e.g.~\cite{chen91}) that a strongly elliptic linear system of second-order differential equations with analytical coefficients has a unique---defined on a compact subset $W\subseteq\Omega$---analytical solution, for smooth Dirichlet boundary data. This establishes the well-posedness of the Dirichlet boundary value problem for the MSE, on sufficiently small regions, provided they are strongly elliptic.

\section{CONDITIONS FOR STRONG ELLIPTICITY OF THE MINIMAL STRAIN EQUATIONS}

\subsection{Derivation of a Strong Ellipticity Condition}

Substituting $D_{j}\rightarrow\xi_{j}$ in $M_{i}^{\; k}$, as given by (\ref{mse2}), we
obtain 
\beq
M_{i}^{\; k}\equiv -\chi^{-1}K_{ij}K^{lk}\xi^{j}\xi_{l}+\fr{1}{2}\left(
\xi_{i}\xi^{k}+\delta_{i}^{k}\xi_{j}\xi^{j} \right).
\eeq
When a basis has been chosen, this becomes a $(3\times3)$ real matrix. At any given point on the spacelike slice, we can introduce a local orthonormal basis, such that $\gamma_{ij}=\delta_{ij}$. In this basis, $M_{i}^{\; k}$ becomes manifestly symmetric:
\beq
M_{i}^{\; k}=M_{(ik)}=-\chi^{-1}K_{ij}K^{l}_{\;
k}\xi^{j}\xi_{l}+\fr{1}{2}\left(\xi_{i}\xi_{k}+\delta_{ik}|\vec{\xi}|^{2}\right).
\eeq
We can then rotate our basis, so as to diagonalize $K_{ij}$:
\beq
K_{ij}=\mbox{diag}(K_{11},K_{22},K_{33})\equiv\mbox{diag}(k_{1},k_{2},k_{3}).
\eeq
Then $M_{ik}$ takes the form
\beq
M_{ik}\triangleq-\chi^{-1}k_{i}k_{k}\xi_{i}\xi_{k}+\fr{1}{2}\left(\xi_{i}\xi_{k}+\delta_{ik}|\vec{\xi}|^{2}\right),
\eeq
where there is no sum over $i$ and $k$, and the ``$\triangleq$'' denotes equality in the orthonormal basis in which $K_{ij}$ is diagonalized. This can be further simplified by (i) introducing a unit-length vector
field 
\beq
l_{i}\equiv k_{i}\chi^{-1/2},
\eeq
(where ``unit-length'' means $l_{1}^{2}+l_{2}^{2}+l_{3}^{2}=1$) and (ii) noting that $M_{ik}$ is quadratic in $\xi_{i}$, which allows for the normalization
\beq
|\vec{\xi}|^{2}=\sum_{i=1}^{3} \xi_{i}^{2}=1,
\eeq
without loss of generality. Then $M_{ik}$ becomes
\beq
2M_{ik}\triangleq\delta_{ik}+(1-2l_{i}l_{k})\xi_{i}\xi_{k}. \label{mtx} 
\eeq
The problem is now to determine what constraints on the vector field
$\vec{l}$ (the normalized diagonal components of the extrinsic curvature in the orthonormal frame) ensure that the eigenvalues of $M_{ik}$ are positive and finite for arbitrary unit vector fields $\vec{\xi}$.

By inspection, we see that one such constraint is that {\em at most one} of the $l_{i}$ components vanish, as the following counter-example shows: if $\vec{l}=\vec{\xi}\triangleq(1,0,0)$, then
$2M_{ik}\triangleq\mbox{diag}(0,1,1)$. It turns out that this is the necessary and sufficient condition, as we will now show.

Let us consider an arbitrary vector field $\vec{\omega}$ and contract it twice with $M_{ik}$:
\bqa
\sum_{i,k} 2M_{ik}\omega^{i}\omega^{k} & \triangleq &
|\vec{\omega}|^{2}+\sum_{i,k} \left( \xi_{i}\xi_{k}\omega^{i}\omega^{k}-2l_{i}l_{k}\xi_{i}\xi_{k}\omega^{i}\omega^{k}\right) \nonumber \\
& \geq & |\vec{\omega}|^{2}-\sum_{i,k} \xi_{i}\xi_{k}\omega^{i}\omega^{k} \nonumber \\
& \geq & 0,
\eqa
where we have used the fact that $|\vec{\xi}|=|\vec{l}|=1$. Since $\vec{\omega}$ is arbitrary,
this implies that the eigenvalues of $M_{ik}$ are non-negative. The inequality can be made strict if we assume that at least two of the $l_{i}$ components are non-vanishing; then, since $|\vec{l}|^{2}=1$, we have $|l_{i}|<1$ for all $i$, whence
\beq
|l_{i}\xi_{i}|<|\xi_{i}|\Rightarrow |l_{i}l_{j}\xi_{i}\xi_{j}\omega^{i}\omega^{j}|<|\xi_{i}\xi_{j}\omega^{i}\omega^{j}|,
\eeq
and thus 
\beq
\sum_{i,k} M_{ik}\omega^{i}\omega^{k}>0.
\eeq
Therefore, the constraint that at most one of the $l_{i}$ components vanish, is a {\em sufficient} condition for the strong ellipticity of the MSE. To show that this is also a {\em necessary} condition, we need to prove the converse: if the system (\ref{mse}) is strongly elliptic, then at most one of the $l_{i}$ vanishes. Equivalently, we need to prove that, if two of the $l_{i}$ vanish, then (\ref{mse}) is {\em not} strongly elliptic.

Suppose, for definiteness, that $l_{2}=l_{3}=0$, so $l_{1}=\pm1$. Then, choose the unit vector $\vec{\xi}$ to have components $\xi_{1}=1$, $\xi_{2}=\xi_{3}=0$. Equation (\ref{mtx}) then implies that $M_{ik}$ is diagonal with entries $M_{11}=0$, $M_{22}=M_{33}=1$, which has a vanishing eigenvalue, and this implies that the MSE are not strongly elliptic. $\blacksquare$

Summarizing, the necessary and sufficient condition for the strong ellipticity of the MSE is that at most one diagonal component of the extrinsic curvature---evaluated in an orthonormal frame in which $K_{ij}$ is diagonal---vanishes.

\subsection{Algorithm for Testing Strong Ellipticity}

The above strong ellipticity condition is only valid in an orthonormal basis in which the extrinsic curvature is diagonal. Testing this condition for some chosen metric and slicing requires a transformation to such a non-holonomic basis, which can be computationally tedious. It is therefore desirable to have a coordinate invariant algorithm for testing the strong ellipticity condition, which can easily be implemented for a general metric and extrinsic curvature in any basis one wishes. In this subsection we derive such an algorithm. Our algorithm is expressed in terms of the object
\beq
{\cal K}_{ij}\equiv\fr{K_{ij}}{\sqrt{\chi}},
\eeq
with trace
\beq
{\cal K}\equiv{\cal K}_{i}^{i}=\fr{1}{\sqrt{\chi}}K_{i}^{i}.
\eeq
In the orthonormal basis,
\beq
{\cal K}^{2}\triangleq\left(l_{1}+l_{2}+l_{3}\right)^{2}=1+2(l_{1}l_{2}+l_{1}l_{3}+l_{2}l_{3}).
\eeq
Clearly, if ${\cal K}^{2}\neq1$ then at least two of the $l_{i}$ components are non-zero, which implies that the MSE are strongly elliptic. This is a sufficient but not necessary condition, since the MSE can be strongly elliptic {\em and} ${\cal K}^{2}=1$: the equation
\beq
{\cal K}^{2}=1\,\Leftrightarrow\,\sum_{i=1}^{3} l_{i}^{-1} = 0,
\eeq
together with the constraint $\sum_{i=1}^{3} l_{i}^{2}=1$, yields a 1-parameter family of solutions that geometrically correspond to two planes intersecting the unit sphere along two spherical triangles (which are a reflection of each other through the origin), whose sides are great circles. Except for the six corners of those triangles, where two $l_{i}$'s vanish and therefore the MSE fail to be strongly elliptic, along the edges the strong ellipticity condition is always satisfied.

Further specialization can be accomplished by considering the tensors
\beq
{\cal H}_{ij}^{\pm}\equiv{\cal K}_{ik}{\cal K}^{k\;}_{j}\pm{\cal K}_{ij}.
\eeq
In the orthonormal basis,
\beq
{\cal H}_{ij}^{\pm}\triangleq \mbox{diag}(l_{1}^{2},l_{2}^{2},l_{3}^{2})\pm\mbox{diag}(l_{1},l_{2},l_{3}).
\eeq
One of these vanishes if and only if two of the $l_{i}$ components vanish, i.e., if and only if the MSE are not strongly elliptic (SE).

Based on these results, we propose the following simple algorithm to test the strong ellipticity of the MSE in an arbitrary basis:

\vspace{0.2cm}

\noindent $[1]\,\chi=0$ ? Yes $\rightarrow$ \fbox{MSE are {\bf not} SE} No $\rightarrow$ [2]

\vspace{0.2cm}

\noindent $[2]\,{\cal K}^{2}=1$ ? Yes $\rightarrow$ [3]. No $\rightarrow$ \fbox{MSE {\bf are} SE} 

\vspace{0.2cm}

\noindent $[3]\,\mbox{Is one of} \,{\cal H}_{ij}^{\pm}=0$ ? Yes $\rightarrow$ \fbox{MSE are {\bf not} SE} No $\rightarrow$ \fbox{MSE {\bf are} SE} 

\vspace{0.2cm}

We note that an equivalent, coordinate invariant, criterion for strong ellipticity is simply the requirement that at most one of the eigenvalues of $K_{ij}$ vanishes, in {\em any} basis. This follows directly from the fact that a change of basis changes $K_{ij}$ into a similar\footnote{Any two $(n\times n)$ matrices, $A$ and $B$, are said to be similar if there exists a $(n\times n)$ matrix $T$, such that $A=T^{-1}BT$. Similar matrices represent the same linear transformation, in different coordinate systems.} matrix, thereby preserving the eigenvalues.

\section{TESTING THE THORNE TEST-BED METRIC}

Thorne \cite{thorne98} has constructed an analytic metric that he believes is similar, in its geometric and dynamical features, to those for inspiraling, fully relativistic neutron star binaries. He has suggested that this metric be used as a test bed for proposed lapse and shift functions for the IBBH problem. In this section, we ask whether the MSE equations for the lapse and shift are strongly elliptic in the Thorne test-bed metric.

In co-rotating coordinates, $\{T,X,Y,Z\}$, the Thorne metric reads \cite{thorne98}
\beq
{\rm d}s^{2}=-A^{2}{\rm d}T^{2}+\Gamma_{ij}({\rm d}X^{i}+B^{i}{\rm d}T)({\rm d}X^{j}+B^{j}{\rm d}T), \label{thornemetric}
\eeq
where the co-rotating lapse $A$, shift $B^{i}$, and 3-metric $\Gamma_{ij}$, are explicit functions of the
spatial coordinates $\{X,Y,Z\}$, the two stars' initial mass $m$ and their time-evolution separation $a(T)$; $a(T)$ is freely adjustable but for approximate radiation-reaction-driven inspiral can be chosen to be
\beq
a(T) = a_{0}\left(1-\fr{32m^{3}T}{5a_{0}^{4}}\right)^{1/4}.
\eeq
The functions appearing in the metric (\ref{thornemetric}) are:
\bqa
A & = & 1-mF_{+}, \\
F_{\pm} & \equiv & \left[ (X-a)^{2}+Y^{2}+Z^{2}+b^{2}\right]^{-1/2} \nonumber \\
&& \pm \left[(X+a)^{2}+Y^{2}+Z^{2}+b^{2}\right]^{-1/2}, \\
\vec{B} & = & \Omega Y(P-1)\partial_{X} \nonumber \\
&& + \left[ X-P\left( X+\fr{4a^{3}F_{-}}{\sqrt{R^{2}+a^{2}}} \right) \right] \Omega \partial_{Y}, \\
P & \equiv & \fr{4ma^{2}}{(R^{2}+a^{2})^{3/2}}, \\
\Gamma_{ij} & = & \Gamma_{ij}^{C}+\Gamma_{ij}^{TT}, \\
\Gamma_{ij}^{C} & = & \delta_{ij}(1+mF_{+})^{2}, \\
\Gamma_{ij}^{TT}, & = & h_{+}(\hat{\theta}\otimes\hat{\theta}-\hat{\phi}\otimes\hat{\phi})+h_{\times}(\hat{\theta}\otimes\hat{\phi}+\hat{\phi}\otimes\hat{\theta}), \\
h_{+} & = & \fr{-4ma^{2}\Omega^{2}}{R}(1+\cos^{2}\theta )\cos [2\bar{\phi}-2\Omega(T-R)], \\
h_{\times} & = & \fr{-4ma^{2}\Omega^{2}}{R}2\cos\theta[2\bar{\phi}-2\Omega(T-R)], \\
\hat{\theta} & = & \fr{\partial_{\theta}}{R}=-\fr{\sqrt{R^{2}-Z^{2}}}{R}\partial_{Z} \nonumber \\ 
&& +\fr{Z}{R\sqrt{R^{2}-Z^{2}}}(X\partial_{Y}-Y\partial_{X}), \\
\hat{\phi} & = & \fr{\partial_{\phi}}{\sqrt{R^{2}-Z^{2}}}=\fr{1}{R^{2}-Z^{2}}(X\partial_{Y}-Y\partial_{X}), \\
\bar{\phi} & = & \phi+\int \Omega{\rm d}T\simeq\phi+\Omega T.
\eqa
Brady, Creighton and Thorne \cite{bct98} (BCT) have shown that for any metric (including Thorne's test-bed) that has the form appropriate to a slowly inspiraling binary, there exists a solution to the MSE that keeps the coordinates co-rotating to within an accuracy $\sim\tau_{\rm rot}/\tau_{\rm rr}\sim(\Omega\tau_{\rm rr})^{-1}\ll1$. If the MSE are strongly elliptic, and if appropriate boundary conditions are imposed on them, then that BCT solution should be their unique solution. The key issue of strong ellipticity should then be tested in the Thorne metric for a slicing that is close to that of the BCT solution. The natural such slicing is one of constant coordinate time $t=T$, since Thorne's coordinates are co-rotating; and for that slicing the extrinsic curvature is
\beq
K_{ij}=\fr{1}{2A}\left[ -\fr{\partial}{\partial T}\gamma_{ij}+2D_{(i}B_{j)}\right]_{T=T_{0}}.
\eeq

We have performed such a test, using our algorithm (end of Sec. 3), for the parameter set $\{M=1.0, b=2.5, a_{0}=4.0\}$. Due to the symmetries of the problem, only one octant of ${\Bbb R}^{3}$, locally defined by the cartesian basis $\{\partial_{X},\partial_{Y},\partial_{Z}\}$, needed to be considered; we chose the region ${\cal C}^{3}:0\leq X,Y,Z\leq10$ and used a grid-size interval of $10^{-3}$. 

We found that $\chi^{2}\neq0$ for all the points tested, but ${\cal K}^{2}=1$ for a set of points topologically equivalent to a 2-sphere with equator on the $X-Y$ plane: for a given value of $Z$, the 2-surface given by ${\cal K}^{2}(X,Y)-1$ intersects the $X-Y$ plane along a topological circle, whose radius decreases with increasing values of $|Z|$. For {\em that} set of points, ${\cal H}_{ij}^{\pm}\neq0$, and thus the MSE are strongly elliptic. The cube ${\cal C}^{3}$ extends sufficiently far from the two masses, that $K_{ij}$ and the other tensors built from it have the same behavior as at asymptotic spatial infinity; in particular, the relevant scalar quantities do not change sign.

\subsection{Generality of the MSE's Strong Ellipticity}

We have shown that the MSE are strongly elliptic in a particular test-bed metric, which is an approximation to that of a relativistic inspiraling binary. It is reasonable to expect that for {\em any realistic} metric for inspiraling binaries, the MSE will also be strongly elliptic, since the failure of strong ellipticity requires a highly non-generic situation---at least two of the diagonal elements of $K_{ij}$ must vanish simultaneously in an orthonormal frame in which $K_{ij}$ is diagonal. Such non-genericity is likely to result only from a very great spacetime and slicing symmetry and/or very special algebraic properties of the metric functions.

A trivial example that illustrates the connection between the failure of strong ellipticity of the MSE and a high degree of symmetry is the Schwarzschild metric: 
\bqa
{\rm d}s^{2} & = & -f(r){\rm d}t^{2}+f^{-1}(r){\rm d}r^{2}+r^{2}({\rm d}\theta^{2}+\sin^{2}\theta {\rm d}\phi^{2}), \\
f(r) & \equiv & 1-2Mr^{-1}.
\eqa
With the usual Schwarzschild slicing (surface of constant time $t$), the MSE are {\em not} strongly elliptic:
\beq
K_{ij}=-\fr{1}{2\sqrt{f(r)}}\partial_{t}\gamma_{ij}=0.
\eeq
In fact, for {\em any} static metric, $\partial_{t}g_{ab}=0$ with slicing $t=$ const., the extrinsic curvature vanishes identically, and the MSE are not strongly elliptic [{\em cf.} Eq. (\ref{mse2})]. On the other hand, if we choose a Lema\^{\i}tre slicing of Schwarzschild (corresponding to the reference frame of radially infalling observers):
\bqa
{\rm d}s^{2} & = & -{\rm d}\tilde{t}^{2}+\fr{2M}{r}\,{\rm d}\tilde{r}^{2}+r^{2}({\rm d}\theta^{2}+\sin^{2}\theta {\rm d}\phi^{2}), \\
{\rm d}\tilde{t} & \equiv & {\rm d}t+\sqrt{2Mr^{-1}}f^{-1}(r)\,{\rm d}r, \\
{\rm d}\tilde{r} & \equiv & {\rm d}t+\sqrt{r/(2M)}f^{-1}(r)\,{\rm d}r, \\
r(\tilde{r},\tilde{t}) & = & \left[ \fr{3}{2}\sqrt{2M}(\tilde{r}-\tilde{t})\right]^{2/3},
\eqa
then the MSE {\em are} strongly elliptic. In particular, 
\beq
K_{ij}=-\fr{1}{2}\partial_{\tilde{t}}\gamma_{ij}(\tilde{r},\tilde{t})=\sqrt{\fr{2M}{r}}\mbox{diag}(\fr{M}{r^{2}},r,r\sin\theta),
\eeq
is positive definite everywhere except at the central singularity $r=0$, where $K_{\tilde{r}\tilde{r}}$ diverges and the other two diagonal elements vanish. It is trivial to check that $\chi\neq0$ and ${\cal K}^{2}\neq1$, for all points (including $r=0$, where both scalars diverge), thereby verifying strong ellipticity.

In the Thorne test-bed metric, 
which has more symmetries than a generic metric for inspiraling binaries, 
the condition ${\cal K}^{2}\neq1$ (end of Sec. 3) failed to be obeyed only by a set of measure zero, for which the MSE were in fact strongly elliptic (the next necessary and sufficient condition for ellipticity was obeyed). This provides an example of how atypical a spacetime point would need to be in order for the MSE not to be strongly elliptic, even in an idealized metric. 

Based on the above observations, we conjecture that in a {\em generic} metric for inspiraling binaries, with a slicing near that which makes $g_{ab}$ evolve on the timescale $\tau_{\rm rr}$, the MSE will be strongly elliptic, thence leading to a unique analytical solution for the choice of lapse and shift.

\subsection{A Note on the Nature of the Boundary Data}

On every spacelike slice, $\Sigma_{t}$, appropriate boundary conditions for $\beta_{i}$ need to be specified on an outer boundary $\partial\Sigma_{t}$ that contains the two horizons, $H_{1}$ and $H_{2}$, in its interior, and on these two inner boundaries, $H_{1}$ and $H_{2}$. Provided $\partial\Sigma_{t}$ is sufficiently far from the two holes, we can take $\beta_{i}$ on $\partial\Sigma_{t}$ to be the co-rotating one at asymptotic spatial infinity (e.g. like the one contained in Thorne's test-bed metric \cite{thorne98}), which is a Dirichlet-type boundary condition.

The inner boundaries are strong-field regions, and in general Neumann-type boundary conditions are needed \cite{smarr&york78a,smarr&york78b} to ensure that $\beta_{i}$ is indeed a minimal distortion one, which will typically require $\beta_{i}\neq0$ on $H_{1}$ and $H_{2}$. Fixing the value of $\beta_{i}$ itself, rather than its normal derivative on the inner boundaries, might lead to the pilling up of distortions near the boundary \cite{smarr&york78b}. Unlike the Dirichlet problem, the Neumann problem does not admit a unique solution, even in an arbitrarily small compact set [{\em cf.} eqs. (8)-(9); we can always add to $u^{k}$ any function that is constant on each connected component of $\Omega$], which precludes the {\em a priori} success of the minimal strain prescription for such type of boundary data.

While this is still an open issue that deserves further investigation (a detailed understanding of how the metric evolves near the horizons through the IBBH phase), it may be that Dirichlet-type data may actually succeed in providing realistic and numerically stable boundary data on the inner boundaries. As long as adiabaticity is maintained, the shift generated by the MSE is such that the coordinates are kept co-rotating and the metric components evolving on the slow gravitational radiation reaction timescale, even {\em near} the horizons \cite{bct98}. It is then reasonable to expect the normal (`time') derivative of the shift near the horizons to evolve rather slowly. In such case, we could then compute the inner boundary values of $\beta_{i}$ on $\Sigma_{t}$ from those on $\Sigma_{t-\Delta t}$ (e.g. by changing them linearly), thereby avoiding having to solve a Neumann-type problem for the inner boundaries at every spacelike slice. For the initial slice, $\Sigma_{t=0}$, we could take $\beta_{i}$ on the inner boundaries to be the co-rotating one, constructed from a post-Newtonian approximation for the early inspiral phase. We must stress that this is a rather naive argument, and that only a detailed numerical study of the spacetime evolution near the horizons, through the IBBH phase, can resolve this question unequivocally.

\section{Conclusions}

The BCT proposal \cite{bct98} for the optimal choice of lapse and shift in a $3+1$ numerical relativity scheme is encoded in a coupled set of second-order partial differential equations: the MSE. We have shown that the MSE are strongly elliptic if and only if at most one of the diagonal components of the extrinsic curvature, evaluated in an orthonormal frame in which the extrinsic curvature is diagonal, vanishes; or, equivalently, if and only if the extrinsic curvature $K_{ij}$ possesses at most one zero eigenvalue. An equivalent coordinate invariant algorithm was derived for general applicability to arbitrary choice of basis and slicing.

We tested this algorithm in the Thorne test-bed metric \cite{thorne98} and found the MSE to be strongly elliptic. The failure of strong ellipticity requires extremely atypical situations, normally associated with highly symmetric metrics and/or special algebraic properties of the metric functions, which are very unlikely to occur in any realistic metric for inspiraling binaries. We therefore expect the MSE to be strongly elliptic for a {\em generic} metric for inspiraling binaries.

Evolving the IBBH spacetime in a $3+1$ numerical scheme, using the minimal strain prescription, will typically require solving a hyperbolic system for $\gamma_{ij}$ and $K_{ij}$, evolving forward in some time $t$, together with an elliptic system for the lapse $\alpha$ and shift $\beta_{i}$, at each surface of constant $t$, $\Sigma_{t}$. The main steps of such a computation can be outlined as follows: (i) specify initial values for $\gamma_{ij}$ and $K_{ij}$; (ii) solve the elliptic MSE system (\ref{mse}), using appropriate boundary values for $\beta_{i}$, to obtain $\beta_{i}$ everywhere on that spacelike slice; (iii) solve Eq. (\ref{lapse}) for the lapse $\alpha$, using the $\beta_{i}$ obtained from the MSE; (iv) solve the dynamical (hyperbolic) equations for $\gamma_{ij}$ and $K_{ij}$ using $\alpha$ and $\beta_{i}$ produced by the MSE. This fully determines the spacetime geometry at the next spacelike slice, $\Sigma_{t+\Delta t}$; (v) repeat steps (ii)-(iv), until the gravitational radiation reaction becomes too large for the evolution to be adiabatic. A simple, computationally inexpensive, self-consistency test would be to check that $||\gamma_{ij}(t)/\gamma_{ij}(t-\Delta t)|-1|\sim(\Omega\tau_{\rm rr})^{-1}\ll1$, for a given set of points at every spacelike slice.

\section*{ACKNOWLEDGEMENTS}

I am grateful to Kip Thorne for invaluable advice, and to Patrick Brady, Jolien Creighton and James York for useful discussions. This work was supported by
F.C.T. (Portugal) Grant PRAXIS XXI-BPD-16301-98, and by NSF Grant PHY-9900776 and NASA Grant NAG5-6840.


\begin{thebibliography}{99}

\bibitem{bct98}
P. R. Brady, J. D. Creighton and K. S. Thorne, Phys. Rev. D {\bf 58}, 061501 (1998)

\bibitem{thorne98}
K. S. Thorne, {\it ``Numerical Relativity for Inspiraling Binaries in Co-Rotating Coordinates: Test Bed for Lapse and Shift Equations''}, gr-qc/980824, {\it Phys. Rev. D} submitted

\bibitem{giaquinta83}
M. Giaquinta, {\it Multiple Integrals in the Calculus of Variations and Nonlinear Elliptic Systems}, (Ann. Math. Studies, {\bf 105}, Princeton University Press: Princeton, 1983)

\bibitem{chen91}
Ya-Z. Chen and Lan-C. Wu, {\it Second Order Elliptic Equations and Elliptic Systems}, (Mathematical Monographs, {\bf 174}, AMS Publishing, 1998), pp. 131-135

\bibitem{smarr&york78a}
L. Smarr and J. W. York, Phys. Rev. D {\bf 17}, 1945 (1978)

\bibitem{smarr&york78b}
L. Smarr and J. W. York, Phys. Rev. D {\bf 17}, 2529 (1978)

\end{thebibliography}
\end{document}